\documentclass[11pt]{article}
\usepackage{hyperref}
\usepackage{graphicx}
\usepackage{amsmath}
\usepackage{amssymb}
\usepackage{epstopdf}
\usepackage{natbib}
\usepackage{color}
 \usepackage{float}
 \usepackage[version=3]{mhchem}

\pdfoutput=1
\begin{document}
\title{Squeezing through: capsule or bubble?}
\author{Geoffrey Dawson and Anne Juel \\
\\\vspace{6pt}
 Manchester Centre for Nonlinear Dynamics and School of Mathematics, \\ University of Manchester, Manchester M13 9PL, UK}
\maketitle
%% The abstract (in this file, and that submitted as text to arXiv) should include the exact phrase
%% "fluid dynamics video" or "fluid dynamics videos"
\begin{abstract}
In this fluid dynamics video, we compare the deformation of two flexible
particles as they propagate through a sudden constriction of a
liquid-filled channel under constant-flux flow: a gas bubble, and a
capsule formed by encapsulating a liquid droplet in a cross-linked
polymeric membrane. Both bubble and capsule adopt highly contorted
configurations as they squeeze through the constriction, exhibit broadly
similar features over a wide range of flow rates, and rupture for
sufficiently high flow rates. However, at flow rates prior to rupture,
certain features of the deformation allow bubble and capsule to be
distinguished: bubbles exhibit a tip-streaming singularity associated
with critical thinning of the rear of the bubble, while the capsule
membrane wrinkles under large compressive stresses induced by the
constriction.

\end{abstract}
% main text
%\section{Introduction}
%% The format is: \href{URL of video}{name that will appear in the text}

Two identical videos were produced: A high quality
\href{http://www.maths.manchester.ac.uk/~ajuel/DFDvideo2013/}{Video HQ} and a lower quality video
\href{http://www.maths.manchester.ac.uk/~ajuel/DFDvideo2013/}{Video LQ}.

\begin{figure}[h!]
\centering
\includegraphics[trim=3.5cm 4.6cm 3.5cm 3cm, clip=true, width=12cm]{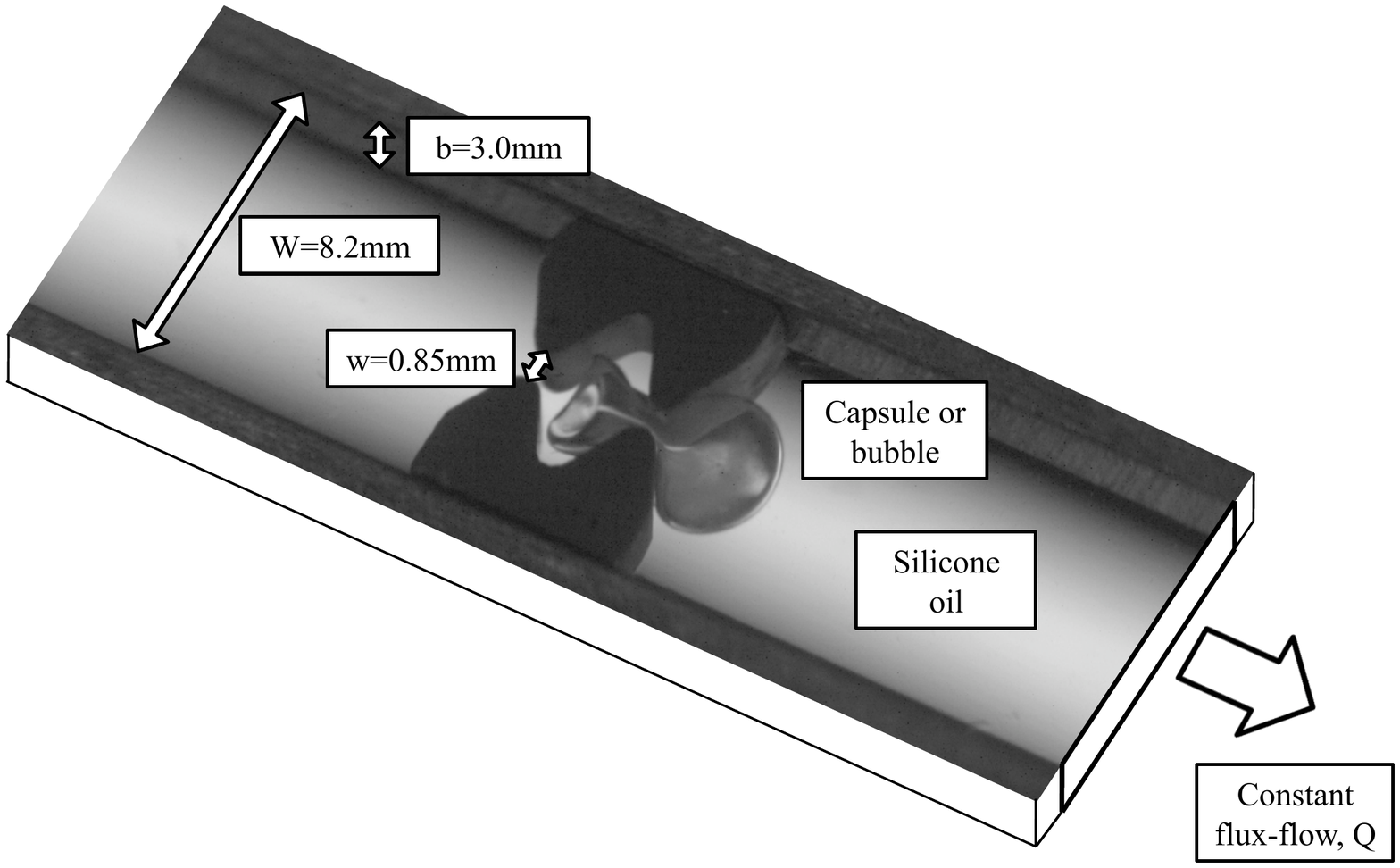}
\caption{A three dimensional view of the experimental setup}
\label{setup}
\end{figure}

The experimental setup is shown in figure \ref{setup}. It consists of a liquid-filled tube of rectangular cross-section, of width $W=8.2$~mm, and depth $b = 3.0$~mm. A local axial variation in width is imposed with a double constriction made from two juxtaposed triangular teeth with rounded ends, each of width $3.25$~mm, so that the minimum width is $w=0.85$~mm. Liquid was withdrawn from the end of the channel at a constant volume-flux, $Q$. Experiments focused on the motion of either a single air bubble or an encapsulated liquid droplet (capsule), which was visualized with a top-view mounted camera.

 The channel was filled with silicone oil with dynamic viscosity $\mu=1.04$~kg~m$^{-1}$~s$^{-1}$, density  $\rho=975$~kg~m$^{-3}$, and surface tension $\sigma=2.1\times 10^{-2}$~Nm$^{-1}$. We used cross-linked albumin alginate capsules, which consisted of a liquid core enclosed by a thin polymeric membrane \footnote{L\'{e}vy, M.-C. and Edwards-L\'{e}vy, F. (2006), J. Microencapsul, 13(2), 169-183}. The capsule had a diameter of $D=3.62\pm 0.16$~mm, a membrane thickness of $0.16\pm0.02$~mm ($4.4\%D$) and a dilation modulus $K=1.0\pm0.1$~Nm$^{-1}$. In all the experiments, inertial forces were negligible (the Reynolds number was $Re=\mu U b /\rho<0.02$, where $U$ is the speed of the bubble/capsule), and the motion was governed by viscous and surface tension forces in the case of bubbles, and viscous and elastic forces in the case of capsules.

In the fluid dynamics video, we compare the motion of a bubble and a capsule squeezing through the double constriction at flow rates chosen to highlight specific phenomena associated with their extreme deformation. At moderate flow rates, the deformation of both bubble and capsule is broadly similar. As the bubble advances through the constriction, its rear tip thins.
This thinning is associated with a divergence of the rear tip speed and
curvature at a critical flow rate. The divergence is suggestive of a
tip-streaming singularity; for flow rates beyond the critical flow rate,
the bubble systematically breaks up. In contrast, the capsule does not exhibit this singularity as the speed of the rear tip always increases linearly with flow rate, and the
capsule ruptures before significant thinning is induced. Moreover, the membrane thickness imposes the value of the maximum curvature locally available to the capsule. However, the capsule membrane can wrinkle as it squeezes through the constriction under sufficient compressive stress imposed by increasing flow rates.

This video demonstrates the propensity of both bubbles and capsules to
undergo extreme geometry-induced deformation, with broadly similar
overall features, but important differences in the detailed deformation
dynamics.

%\begin{enumerate}
%\item An explanation of what is shown in the video.
%\item The relevant conditions, parameters, etc..
%\item References to any papers containing further information on the
%videos.
%\item In the Abstract (in the LaTeX file and in the text submitted
%to arXiv), the exact phrase ``fluid dynamics video" or ``fluid
%dynamics videos". This is to facilitate subsequent searching.
%\end{enumerate}
%
\end{document}